\def\year{2021}\relax

\documentclass[letterpaper]{article} 
\usepackage{aaai21}  
\usepackage{times}  
\usepackage{helvet} 
\usepackage{courier}  
\usepackage[hyphens]{url}  
\usepackage{graphicx} 
\usepackage{natbib}  
\usepackage{caption} 
\usepackage{subfig}
\usepackage{pgfplots}
\pgfplotsset{compat=newest}
\usetikzlibrary{decorations.pathmorphing}
\pgfplotsset{every axis/.append style={
    axis x line=middle,    
    axis y line=middle,    
    axis line style={<->}, 
        xlabel={$x$},          
        ylabel={$y$},          
        label style={font=\small},
        tick label style={font=\small}}}
\tikzset{>=stealth}
\usepackage{multirow}
\title{Applied Machine Learning for Games: A Graduate School Course}
\author{
    Yilei Zeng,
    Aayush Shah,
    Jameson Thai,
    Michael Zyda
    \\
}
\affiliations{
    University of Southern California\\
    \{yilei.zeng, aayushsh, jamesont, zyda\}@usc.edu\\
}

\begin{document}

\maketitle

\begin{abstract}

The game industry is moving into an era where old-style game engines are being replaced by re-engineered systems with embedded machine learning technologies for the operation, analysis and understanding of game play. In this paper, we describe our machine learning course designed for graduate students interested in applying recent advances of deep learning and reinforcement learning towards gaming. This course serves as a bridge to foster interdisciplinary collaboration among graduate schools and does not require prior experience designing or building games. Graduate students enrolled in this course apply different fields of machine learning techniques such as computer vision, natural language processing, computer graphics, human computer interaction, robotics and data analysis to solve open challenges in gaming. Student projects cover use-cases such as training AI-bots in gaming benchmark environments and competitions, understanding human decision patterns in gaming, and creating intelligent non-playable characters or environments to foster engaging gameplay. Projects demos can help students open doors for an industry career, aim for publications, or lay the foundations of a future product. Our students gained hands-on experience in applying state of the art machine learning techniques to solve real-life problems in gaming.



\end{abstract}

\section{Introduction}


Applied machine learning in games is now a vividly expanding research field that provides a platform for novel vision, language, robotics, and online social interaction algorithms. Exposure to state-of-the-art research literature is an integral part of the course plan, in part because research community is moving forward at an ever-increasing speed and understanding several backbone papers will clarify the research question and enhance an understanding of the iterations and improvements made. Moreover, an emphasis on the state-of-the-art research methods fosters an appreciation of research design and methodology, and more generally, of the importance of critical evaluation. Therefore, new ideas can be generated based on critical thinking. 

As this course does not require prerequisites on machine learning, we encourage learning by doing. A self-proposed project will enable the students to tailor themselves into research-oriented, industry-oriented or patent-oriented directions. The projects' difficulties are also dynamically adjustable towards different students' learning curve or prior experiences in machine learning. In this class, we intend to encourage further research into different gaming areas by requiring students to work on a semester-long research project in groups of up to 8. Students work on incorporating deep learning and reinforcement learning techniques in different aspects of game-play creation, simulation, or spectating. 
These projects are completely driven by the student along any direction they wish to explore. Giving the students an intrinsic motivation to engage on their favorite ideas will not only make teaching more time efficient but also bestow a long-term meaning to the course project which will open doors for them. By having a semester-long project, students can dive deep into different algorithms. They also receive hands-on experience incorporating various machine learning algorithms for their use case.

Writing, presenting and teamwork proficiency is a critical component of a higher education, and this courses involve writing assignments, extensive team collaboration and oral presentation to a public audience. Student performance on formal writing assignments, project actualization and public presentation provides benchmarks for examining student progress, both within and across semesters. 

This experience report describes a three semester-long effort in an applied machine learning course with advanced research orientations in gaming. This course has withstood the test through in-person, hybrid learning, and completely online modalities separately. We contribute a new course design inline with the most recent advancements in the gaming research community. This course attracts and caters to mutual interests across engineering graduate programs. 
Of the 292 students enrolled in this course over 3 semesters; 1.3\% major in Environmental Engineering, Physics , Chemistry or Computer Networks, 1.3\% are Software Engineering or High-Performance Computing, 2\% are Game Development, 3.2\% are Electrical Engineering, 4\% are Intelligent Robotics, 7\% are Computer Engineering,  9\% are Applied Data Science,  9.2\% are Data Science and the majority of students, 63\%, are majored in General Computer Science. 
Students are expected to gain both creative and fun hands-on experience through a semester-long applied deep learning and reinforcement learning project. This course demonstrates the feasibility of teaching and conducting state-of-the-art applied machine learning research within mixed focused engineering graduate students. This course also shows the capability to help students open doors for an industry career, aim for publications, or lay the foundations of a future product.


    

\section{Background}
Aiming for approaching Artificial General Intelligence (AGI), video games such as Atari, Doom, Minecraft, Dota 2\footnote{OpenAI Five: \url{https://openai.com/blog/openai-five/} (Last accessed: 12/15/2020)}, StarCraft, and driving games have been used extensively to test the deep learning and reinforcement learning methods' performance and generalizability. Following Google's Alpha Go ~\cite{silver2016mastering}, researchers have made steady progress in improving AI's game playing capabilities. Besides creating intelligent Non-player characters (NPC), game testing and level generation have also seen advancement with deep learning for the gaming industry. Moreover, Machine learning can unleash the power of data generated from millions of players worldwide. Gaming provides numerous behavioral data for online user profiling, advertisement recommendation, modeling social interactions, and understanding decision-making strategies. Apart from in-game trajectories, Esports and streaming open new research opportunities for multi-modal machine learning that combines textual, audio natural language processing, computer vision with social media. Gaming simulated interactive environments can extend beyond gaming and adopt practical values for robotics, health, and broader social good.

We cover all the following topics in our course. The cited work also serve as supplementary reading materials. And these topics will be exemplified in the Student Projects section.

\subsection{Benchmark Environments and Competitions}
For academic and individual researchers, the IEEE Conference on Games(COG), AAAI Conference on Artificial Intelligence and Interactive Digital Entertainment(AIIDE), Conference on the Foundations of Digital Games (FDG), and Conference on Neural Information Processing Systems (NeurIPS) host a series of annual competitions featuring creating deep learning and reinforcement learning algorithms for game-play generation or AI playing games.

Major technology companies open-sourced a number of gaming AI environments to help push forward the boundaries of Artificial general intelligence (AGI). OpenAI releases for public OpenAI Gym ~\cite{brockman2016openai}, which incorporates Arcade Learning Environment (ALE) that emulates Atari 2600 game-playing ~\cite{bellemare2013arcade}, robotics, and expanding third party environments. Gym Retro ~\cite{nichol2018retro} extends the integration to 1000 retro games, including games from the Sega Genesis and Sega Master System, and Nintendo’s NES, SNES, and Game Boy consoles. Facebook AI has released ELF: An Extensive, Lightweight, and Flexible Platform for Game Research ~\cite{tian2017elf}, which provides three environments, i.e., MiniRTS, Capture the Flag, and Tower Defense. PySC2 is DeepMind's Python component of the StarCraft II Learning Environment (SC2LE) ~\cite{vinyals2017starcraft}. STARDATA ~\cite{lin2017stardata}, a StarCraft: Brood War replay dataset, is published with the StarCraft II API. Microsoft announced Project Malmo ~\cite{johnson2016malmo}, which provides an open-source platform built on top of Minecraft. MineRL Environments built on Malmo are released for NeurIPS competitions and MineRL imitation learning datasets~\cite{johnson2016malmo} with over 60 million frames of recorded human player data are published to facilitate research. The Unity Machine Learning Agents Toolkit (ML-Agents) ~\cite{juliani2018unity} is an open-source project that enables games and simulations created by individuals to serve as environments for training intelligent agents. As an active research field, new environments and tasks emerge daily. We leave the constant learning to students as they progress through their projects.

\subsection{Computer Vision \& Natural Language Processing} 
    
Learning to play from pixels have become a widely accepted approach for traning AI agents after DeepMinds paper of playing Atari with Deep Reinforcement Learning ~\cite{mnih2013playing} using raw pixels as input. Vision-based user inputs augmented automatic face, and gesture recognition has enabled the fitness game genre to boost. With the pandemic in 2020, virtual reality devices and fitness gaming has offered a safe and entertaining indoor option. With the booming of streaming platforms, elaborate walk-through, strategies, and sentiments shared via videos provided a wealth of data for applied computer vision tasks such as motion analysis and activity recognition. Leveraging the information provided in the YouTube videos, researchers can guide deep reinforcement learning explorations for games with sparse rewards~\cite{aytar2018playing}. 
    
Understanding players' textual interactions, both in-game and on social networks, is crucial for gaming companies to prevent toxicity and increase inclusion. In gaming, language generation techniques are leveraged to generate narratives for interactive and creative storytelling. Text adventure games is an active task for reinforcement learning (RL) focused Natural Language Processing (NLP) researchers. Microsoft introduced TextWorld ~\cite{cote18textworld}, a text-based game generator, as send box learning environment for training and testing RL agents.

Recent progress on deep representations on both computer vision and natural language processing have enabled  the exploration on issues of active perception, long-term planning, learning from interaction, and holding a dialog grounded in an simulated environment. Simulated housing environments such as the ALFRED (Action Learning From Realistic Environments and Directives) ~\cite{ALFRED20} project in Allen Institute and Habitat Lab from Facebook research ~\cite{habitat19iccv}, serve for embodied AI tasks (e.g. navigation, instruction following, question answering), configuring embodied agents (physical form, sensors, capabilities), training these agents (via imitation or reinforcement learning), and benchmarking their performance on the defined tasks using standard metrics. AI and language instructed MiniRTS project ~\cite{hu2019hierarchical} from Facebook AI is similar to this initiative.

\subsection{Player Modeling and Human AI Interactions}
Social gaming, such as the Battle-Royale genre and Animal Crossing, has gained increasing popularity. Combined with heterogeneous data provided on social media and streaming platforms, understanding and predicting players' behavior patterns considering graph structures becomes increasingly important. The data provided by major AAA games will offer resources to imitating and modeling human behaviors ~\cite{sapienza2018individual,zeng2020how} and facilitate understanding of human collaborations~\cite{zeng2019influence}.

Gaming industry with exuberant data of in-game human collaborations makes suitable sand-box environments for conducting multi-agent interaction/collaboration research. For instance, multi-agent Hide-and-Seek ~\cite{baker2019emergent}, OpenAI Five ~\cite{berner2019dota}, AlphaStar ~\cite{vinyals2019grandmaster}, Hanabi ~\cite{bard2020hanabi} and capture the flag ~\cite{jaderberg2019human} are some initial attempts.

With detailed human behavior trajectory recorded as replays or demos, gaming environments provide data-intensive sources for human-computer interaction research. Recent advancements of AI in games has evolved human-computer interactions in gaming environments into human bots interactions. As suggested in paper ~\cite{risi2020chess}, with the increasing popularity in human/AI interactions, we will see more research on human-like NPC and human-AI collaboration in the future.

\subsection{Procedural Content Generation}    
Procedural Content Generation via Machine Learning (abbreviated PCGML) ~\cite{summerville2018procedural} embraces a broadening scope, incorporating automatic generation of levels, gaming environments, characters, stories, music, even game-play mechanics. In the future, more reliable and explainable machine learning algorithms will emerge in this direction.

\subsection{Simulated Interactive Environments and beyond}

Playtesting, matchmaking, dynamic difficulty adaptation (DDA) are some other important tasks for gaming industry to solve using machine learning.

Beyond gaming, interactive environments are used to mimic real-life scenes such as training robots or autonomous vehicles. Interactive gaming environments can also serve as demonstrations for game theory decision makings that serve AI for social good initiatives.

\section{Course Design}

The semester-long course comprises 15 lectures. The detailed course structure consists of weekly lectures on deep learning and reinforcement learning fundamentals, project demonstrations of how each technique are applied in gaming use cases and openly available tools or environments. Upon the conclusion of the lecture, each team updates their weekly progress to the course instructors. Every alternate week students conduct a power-point presentation along with a demo on their team’s progress to the entire class. We encourage the students to be prepared with questions before class to learn proactively rather than learning passively. The instructor evaluates the progress and provides either algorithmic suggestions or structural suggestions to facilitate their learning and project formulation every week. 

We host the midterm and final on the 8th and 15th week. Each team will present PowerPoint and live or recorded demos on their project on both midterm and final. We will also collect the Engineering Design Document (EDD) and a technical paper draft on both midterm and final to foster continuous contribution. We require each team to construct a website to present their project demos to help them on the job market. The gradings' weights are 20\% for mid-term EDD, 20\% for the mid-term draft of technical paper, 10\% for midterm presentation, 20\% for final EDD, 20\% for final of technical paper, 10\% for final presentation.

The learning objective for the course:
(1) Students learn deep learning and reinforcement learning fundamentals through lectures and supplemental materials;
(2) Students learn the most recent advancements, landscape, and applied use cases of machine learning for gaming;
(3) Students can unleash their creativity in projects that cater to their career plans; 
(4) Students engage in teamwork and practice both oral and written presentation skills.

The course first introduces students to the difference between Artificial Intelligence, Machine Learning, and Deep Learning \cite{8259629}. We then cover the survey of Deep learning applications in games \cite{8632747} to give students a tentative idea on projects they can pursue. Following the lecture, students must select a machine learning project and the game they will work on. The course instructors will guide and instruct students' projects according to the sub-directions shown in backgrounds, i.e., benchmark environments and competitions, computer vision and natural language processing, player modeling and human-AI interactions, procedural content generation, simulated interactive environments, etc.

Apart from building a new research project from scratch, students can choose to advance on projects created in the previous semesters for better algorithmic AI performances.

In the first half of the course, we introduce the fundamentals of deep learning. We start with the concept of backpropagation \cite{hecht1992theory}, along with gradient descent \cite{baldi1995gradient} is covered to solidify student’s theoretical understanding of Neural Networks. The different activation functions covered include the sigmoid, tanh \cite{lecun2012efficient} and ReLu \cite{nair2010rectified} functions. We cover a tutorial on combining Neural Networks with Genetic Algorithms in a simulated game environment for Flappy Bird. Students are then introduced to popular Deep Learning frameworks like Tensorflow and Pytorch. 

We then move onto Convolutional Neural Networks (CNNs). Students are introduced to the convolution layer, pooling layer, and fully connected layer along with their respective functionalities. We also cover appropriate activation functions and loss functions for CNNs. A brief overview of state-of-art deep CNN based architectures for object detection tasks are given to students. These include R-CNN \cite{girshick2014rich}, Fast R-CNN \cite{girshick2015fast}, Faster R-CNN \cite{ren2015faster} and YOLO \cite{redmon2016you,redmon2017yolo9000,redmon2018yolov3}. We cover a sample program on image classification tasks \cite{lee2018image} using Tensorflow. Students are encouraged to experiment with the source code and try different CNN configurations to improve the classifier’s accuracy.

Following CNN, we explore different variants of a Recurrent Neural Network (RNN) \cite{graves2012supervised}. RNNs are used for sequence tasks. Long short-term memory (LSTM) \cite{hochreiter1997long} overcome the exploding and vanishing gradient problems \cite{hochreiter1998vanishing,pascanu2012understanding} in vanilla RNN, which enables them to learn long term dependencies more effectively. We explore a case study on LSTM-based architecture implemented for the game of FIFA 18.\footnote{FIFA 18 AI (Last accessed: 12/15/2020): \url{https://github.com/ChintanTrivedi/DeepGamingAI_FIFA} }After 400 minutes of training, the LSTM based bot scored 4 goals in 6 games of FIFA 18 on beginner difficulty.

Moving on, we introduce Generative Adversarial Networks (GANs) \cite{goodfellow2014generative} and its variations. We then give an example of using GANs to generate high-quality anime characters \cite{jin2017towards}.

In the second half of the course, we introduce the fundamentals of reinforcement learning. We start by answering the following questions: What is Reinforcement Learning? Why is it needed in games? What are its advantages in games? Why can’t we use supervised learning in games? We then introduce Markov Decision Process (MDP), Partially Observable Markov Decision Process (POMDP) \cite{mnih2015human, astrom1965optimal}, value iteration \cite{bellman1957markovian} and policy iteration. 
 
We move on to introduce Q-learning \cite{watkins1989learning} and Deep Q-Networks (DQN) \cite{mnih2013playing}. In 2013, a Deep Q-Network was applied to play seven Atari 2600 games \cite{mnih2013playing}. In 2015 the same network was used to beat human-level performance in 49 games \cite{mnih2015human}. For this course we ask students to refer to a sample program that uses a DQN for Flappy Bird game \footnote{Flappy Bird AI (Last accessed: 12/15/2020): \url{https://yanpanlau.github.io/2016/07/10/FlappyBird-Keras.html}}. Students are encouraged to tune the model's parameters and run the training scripts to get a better practical understanding of Deep Q-Learning.  

Lastly, we introduce students to Policy Gradient algorithms \cite{kakade2002natural}. Policy gradient based algorithms such as Actor-Critic \cite{konda2000actor,fujimoto2018addressing, mnih2016asynchronous} and Proximal Policy Optimization \cite{schulman2017proximal} have provided state of art performance for Reinforcement Learning tasks \cite{stooke2018accelerated}. A Case Analysis to play Torc, a racing car game, using Policy Gradient is covered \footnote{TORCS AI (Last accessed: 12/15/2020): \url{https://yanpanlau.github.io/2016/10/11/Torcs-Keras.html} } to supplement the material covered in class. Students are given a chance to develop their agents to play the game Dino Run \footnote{Dino Run AI (Last accessed: 12/15/2020): \url{https://blog.paperspace.com/dino-run/} } and compete with the remainder of the class.

\section{Reading Assignments}

Material for reading assignments primarily stems from Andrew Glassner's textbook titled Deep Learning: From Basics to Practice. This course is supplemented by various sources, including articles on websites such as Medium, TowardsDataScience, tutorials from GDC, TensorFlow, Pytorch, OpenAI Gym, ML-Agents, and survey papers of recent advancements in gaming AI research. These materials incorporate detailed information on implementing specific deep learning or reinforcement learning algorithms, step-by-step guides for implementing a gaming AI project from scratch, and state-of-the-art research papers as references.

\section{Guest Lectures}

We invited 2-3 guest lecturers every semester who were either experienced professionals from the gaming industry or Ph.D. students researching Deep Learning and Reinforcement Learning for games. These lecturers provided valuable insights to students into how machine learning is applied in different gaming research areas. Some of the topics covered in these lectures include applications of Deep Learning in Zynga, Policy Gradient based agents for Doom, and current research frontiers for machine learning in gaming. The lecturers also attended student presentations and provided students with feedback on technologies that they could utilize for their respective projects. 

\begin{figure}[ht]
\includegraphics[width=8cm]{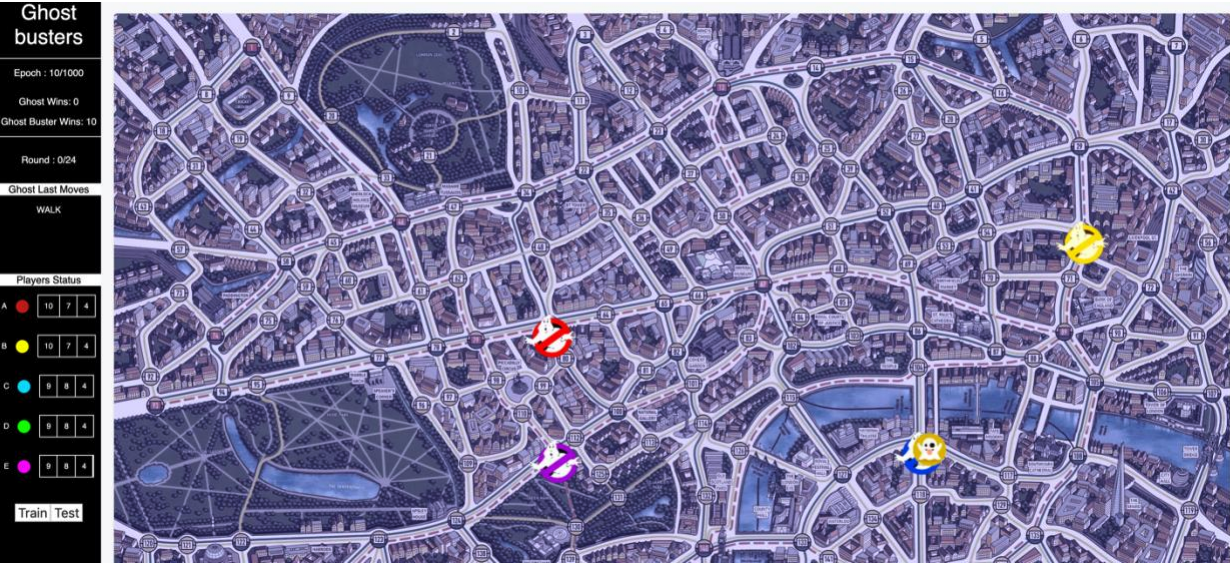}
\caption{DQN based agent for Ghostbusters}
\label{scotlandyard}
\end{figure}

\section*{Student Projects}
This section selected and summarized 32 student course projects, covering various topics based on the different sub-domains illustrated earlier in the background section. 

\subsection{Machine Learning for Playing Games}
To train AI agents in League of Legends, one project used YOLOv3 object detection algorithm to identify different champions and NPCs in League of Legends. They also trained two separate agents, one combining PPO and LSTM, and one supervised LSTM trained on keyboard and mouse pressed captured from the advanced League of Legends players. In a one-on-one custom game, agents achieved first blood against amateur and average players, respectively \cite{lohokare2020deep}.  

Tackling a tower defense game, one team focused on formulating a strategy to place towers. The agent also had to monitor gold income from destroying monsters and view the best locations and timing to place the tower as well as tower upgrades. Using a CNN, the agent is trained on summarized data of randomly generated tower placements where each sample includes the placement of towers, selling and upgrade of towers, and the last wave number achieved.

Scotland Yard and Ghostbusters are two similar projects that aim to train agents to play hide and seek. The agents use an asymmetric environment with incomplete information to either seek or hide from the other agent. There are one hiding player and five seeker players. For both games, the two teams built DQN based agents with different reward shaping functions for the seekers as well as the hider. Figure~\ref{scotlandyard} shows the environment for training an agent in Scotland Yard.



An agent trained to play the online multiplayer game Slither.IO aim to achieving a high score against other players. Applying a DQN and Epsilon Greedy Learning Strategy, the agent observed the game's current frame to determine a direction to move in.

PokemonShowdown is an online game simulator to play a one-on-one match of Pokemon. With a predefined Pokemon set, an agent was trained using a DQN with states incorporating the arena state, player active, and reserve states to determine its next actions. Against a minimax agent, the DQN agent won 17 games out of 20 and effectively learned super effective moves and generally avoided minimally effective ones. 

Donkey Kong is a Nintendo 1983 arcade game where Mario has to reach Donkey Kong while dodging barrels. Starting from a minimal interface, a team mapped and fed each object's bit locations to an agent based on a Q-learning. This agent could be further broken down into a priority and background agent. This project successfully produced an agent that can complete the first level in Donkey Kong.

\subsection{Benchmark Environments and Competitions}
MarioKart64 is a benchmark game for numerous tutorials and competitions. Using a CNN and DAGGER algorithm, the team compared their agent's recoverability from going off track or immediately using power-ups. Moreover, the team applied transfer learning to a Donkey Kong racing game. 

Two Pommerman teams worked on building an agent to play the NES game Bomberman. Both teams used Pytorch and TensorFlow but differed in that one focused on PPO and A2C whereas the other team focused on Accelerated Proximal Policy Optimization (APPO). Along with different reward functions, the teams found that PPO and APPO agents on average outperformed the A2C agent in exploring the game board but not necessarily in laying bombs or winning the game.

Inspired by DeepMind's AlphaGo, one team tackled the game of Go with their agent. Despite the hardware difference, the team successfully trained an agent to win over amateur human player. Using greedy and simple neural network agents as a benchmark, the team's agent utilized both traditional and deep learning algorithms to outperform the baseline agents, achieving the same rank as an advanced amateur player (1 dan). 

Another DeepMind inspired team explored the research and underlying architecture of the multi-agent system, AlphaStar, in the RTS environment Starcraft II. Specifically, the project aimed to utilize algorithms such as DQN with experience replay, CNN, Q-learning, and behavior tree to model different agents against an AI. The team successfully trained four agents where each agent played 100 games against an easy AI where the win rates were 13, 68, 96, and 59, respectively.

\begin{figure}[h]
\includegraphics[width=8cm]{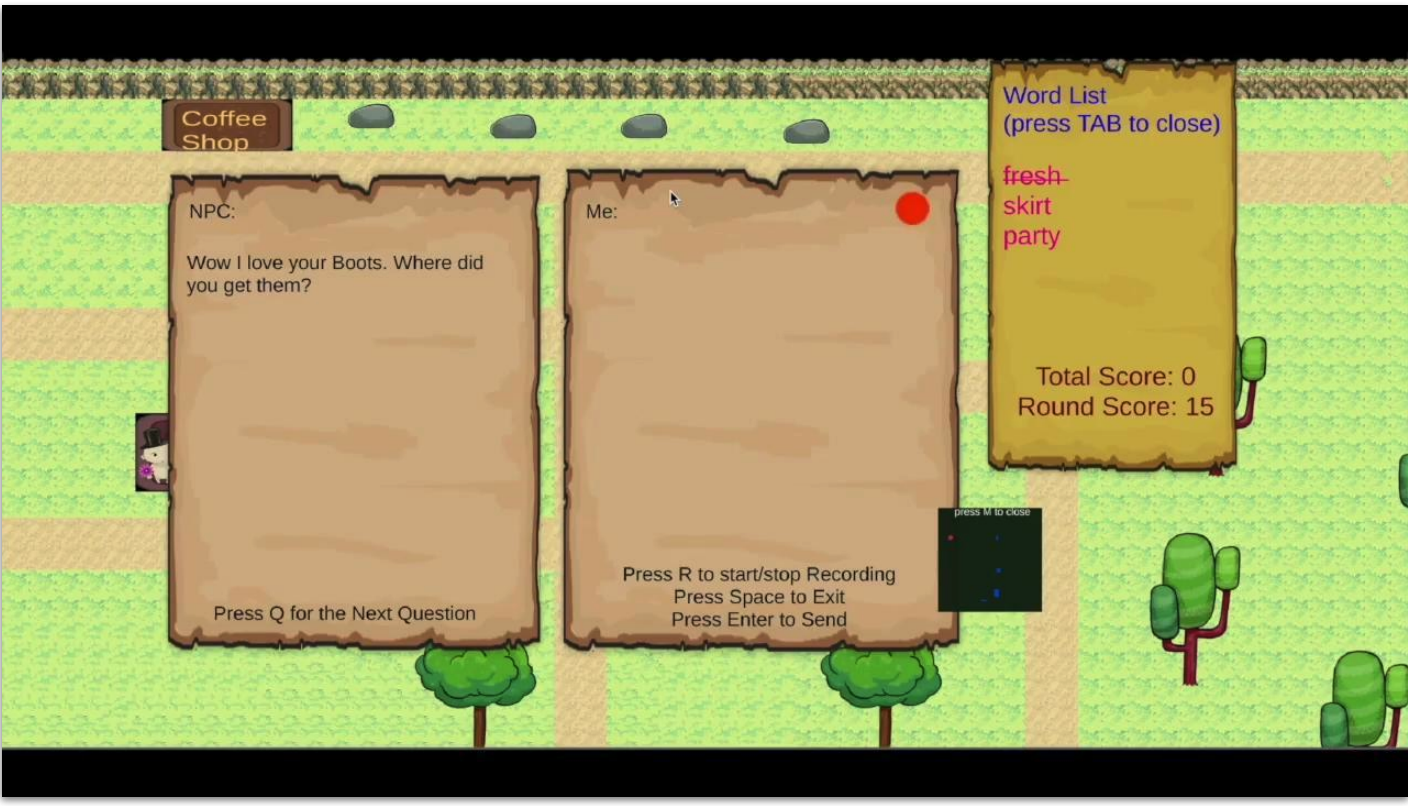}
\caption{Game interface for MapChat: A game designed leveraging text-to-speech  and  automatic  speech  recognition to teach players English}
\label{MapChat}
\end{figure}

\subsection{Computer Vision}

Deep fake applications which uses deep learning to generate fake images or videos have raised debates in AI community. One project applied realistic and personalized head models in a one-shot setting as an overlay to video game characters. They picked Unity3D Mario model for experiment. Taking a video input, the machine learning system extracted facial landmarks on a person and mapped them to a specified character model. In mapping human features to a character Mario, the system primarily looked at detecting the head model as well as specific facial features; the eyes, nose, and mouth.

Counter-Strike Global Offensive (CSGO) is a popular online first-person shooter game. Using object detection models based on Single-Shot Detection and Retinanet, an agent was trained to play the game while distinguishing friends from foes. The agent also demonstrated visual transfer learning between the newer CSGO and the older Counter-Strike 1.6 game, where the model learned low-level features common to both CS 1.6 and CSGO. 


Motion recognition is an important task in computer vision. One team developed a game from scratch while leveraging Computer Vision techniques in Unity 3D. The team created an underwater endless runner game where the agent must overcome random rock hurdles and collect money. Python's OpenCV package was used to detect human body movements and move the submarine correspondingly. As the human player moving left, right, up (jump) or down (crouch), the submarine responded in the same directions via TensorFlow's PoseNet.

A different motion capture project focuses on pose estimation and accurate fidelity for weightlifting form. The team collected data of both good and bad forms of various exercises to be marked and fed into a OpenPose model. They tackled the project in three approaches; splitting the input into a series of periodic frames, summarizing frames, and feeding full frames into a Keras ResNet CNN model. The video is evaluated by a voting system model that tells the user if the exercise had good or bad form. 

\subsection{Natural Language Processing}

Featuring text-to-speech and automatic speech recognition, MapChat helps users practice their English speaking skills through a quest-based role-playing game. Users can move around and complete objectives by speaking prompted phrases in specific locations using a simple map. The audio is recorded and processed to provide feedback to the user regarding how clear and cohesive the response is. Figure \ref{MapChat} shows the game interface developed by students for Mapchat. 

Language generation is a challenging task in NLP. Utilizing FIFA and Pro Evolution Soccer commentaries, a team generated on-the-fly commentary for a football game. Applying NLP techniques, the team fed their Tensorflow model game prompts as seed words to produce relevant sentences that produced coherent and realistic announcer prompts. Another team sourced their information from movie databases like IMDb and IMSDb. With the goal of dialogue generation, their system examines keywords, dialogues, and sentiment from game text. Using multiple models and frameworks such as Markovify, LSTM, and Watson, the team generated coherent character dialogues. 

\subsection{Data Science and Player Modeling}

One CSGO project ~\cite{zeng2020learning} proposes a Sequence Reasoner with Round Attribute Encoder and Multi-Task Decoder to interpret the round-based purchasing decisions' strategies. They adopt few-shot learning to sample multiple rounds in a match and modified model agnostic meta-learning algorithm Reptile for the meta-learning loop. They formulate each round as a multi-task sequence generation problem. The state representations combine action encoder, team encoder, player features, round attribute encoder, and economy encoders to help their agent learn to reason under this specific multi-player round-based scenario. A complete ablation study and comparison with the greedy approach certifies the effectiveness of their model.

Instead of looking at the in-game content of CSGO, another team examined the mental state of the CSGO Twitch audience to detect and define a metric of audience immersion. Representations of different modalities are fed to a multi-modal fusion system. Representations learned through CNN and RNN cover three modalities, i.e., video recordings, commentary audio, and Twitch chat. The model assigns text inputs with positive or negative connotations that later use gameplay audio to capture and map audience immersion.

\begin{table*}[t]

\centering
\resizebox{1.99\columnwidth}{!}{
\begin{tabular}{l|l|l|l|l}
\hline\hline
\begin{tabular}[c]{@{}l@{}}\textbf{Your initial learning motivation}\\ \textbf{for taking this class?} \\ Sample Size 93.\\ 52\% AI in Interactive Entertainment\\ 38\% Computer Vision\\ 29\% Game theory, \\ \hspace{0.75cm}Multi-agent Systems\\ 23\% NLP, Data Science\\ \hspace{0.65cm} Human-Computer Interaction\\ 12\% Procedural Content Generation\\ 5\% \hspace{0.15cm}Gaming Benchmark \\\hspace{0.6cm} Environments and Competitions\end{tabular} & \begin{tabular}[c]{@{}l@{}}\textbf{What have you learned most} \\ \textbf{from this class?} Can choose\\ more than one, sample size 55.\\  \\75\% Hand on Team \\ \hspace{0.8cm}Project Experiences \\ 74\% Applied Machine \\ \hspace{0.8cm}Learning for Gaming \\\hspace{0.8cm}Tasks \\ 53\% Deep Learning and \\ \hspace{0.8cm}Reinforcement Learning \\\hspace{0.8cm}Theory\end{tabular} & \begin{tabular}[c]{@{}l@{}}\textbf{What did you struggle with} \\ \textbf{most in class?}\\ Can choose more than one,\\ sample size 55.\\  38\% Division of Labor \\\hspace{0.75cm}within Team\\ 36\% Applying Deep Learning \\ \hspace{0.75cm}Algorithms\\ 31\% Finding the topic\\ 16\% Finding the team\\ 9\% \hspace{0.15cm}Lost during \\ \hspace{0.75cm}weekly progress\end{tabular} & \begin{tabular}[c]{@{}l@{}}\textbf{On a scale of 1(low)}\\ \textbf{-5(high), how  would}\\ \textbf{you recommend}\\  \textbf{this class to your} \\\textbf{friends?} \\Sample size 55.\\ \\  0\%  \hspace{0.9cm}1\\  2\% \hspace{0.9cm}2\\ 9\% \hspace{0.9cm}3\\ 29\% \hspace{0.75cm}4\\ 60\% \hspace{0.75cm}5\end{tabular} & \begin{tabular}[c]{@{}l@{}}\textbf{On a scale of 1(low)}\\ \textbf{-5(high), how do you}\\  \textbf{think this class will} \\ \textbf{help you find a} \\ \textbf{full-time job or} \\ \textbf{internship?}\\ Sample size 55.\\  0\% \hspace{0.9cm}1\\ 5\% \hspace{0.9cm}2\hspace{0.8cm}\\ 33\% \hspace{0.75cm}3\\ 51\% \hspace{0.75cm}4\\ 11\% \hspace{0.75cm}5\end{tabular} \\ \hline \hline
\end{tabular}}
\caption{Our evaluation for the class is based on two sets of surveys containing five questions in total. The detailed survey statistics are listed below each question.}
\label{tab:evaluation}
\end{table*}

\subsection{Procedural Content Generation}
Part of the famous Mario genre, Super Mario Bros. is a side-scrolling game where levels are meticulously designed from scratch. However, with procedural generation, a level can be produced and deployed with minimal or no design changes. Using an RNN, LSTM, and Markov Chain model, the team map a sequence of characters to an object in the Mario world that is later interpreted as a Mario Level. Each generated level is evaluated by an A* agent to determine if the agent can complete the level. Ultimately the Markov model produced the best ratio of completed to incomplete levels followed by the RNN and LSTM models.

Generating creative arts play an important role in gaming. One team worked on constructing a GAN for character and asset creation in their custom-Unity game. In addition to building a GANs, the team used Unity's ML-agents libraries as a framework to build offensive and defensive AI's which were trained with different reward functions. 

Using conditional GANs, one team augmented real videos with stylized game environments. A reference style image is used as an input to encode two vectors to generate a Gaussian random noise based model to a video generator. SPADE ResNet blocks are then used to reinforce the segmentation mask and provide coherent and consistent video frames. The constructed standalone model could be used to apply a reference style to any real-world input video.

\begin{figure}[ht]
    \centering
    \subfloat[Quad-copter Project]{\label{fig:quadcopter}
        \includegraphics[width=3.8cm, height = 3.5cm]{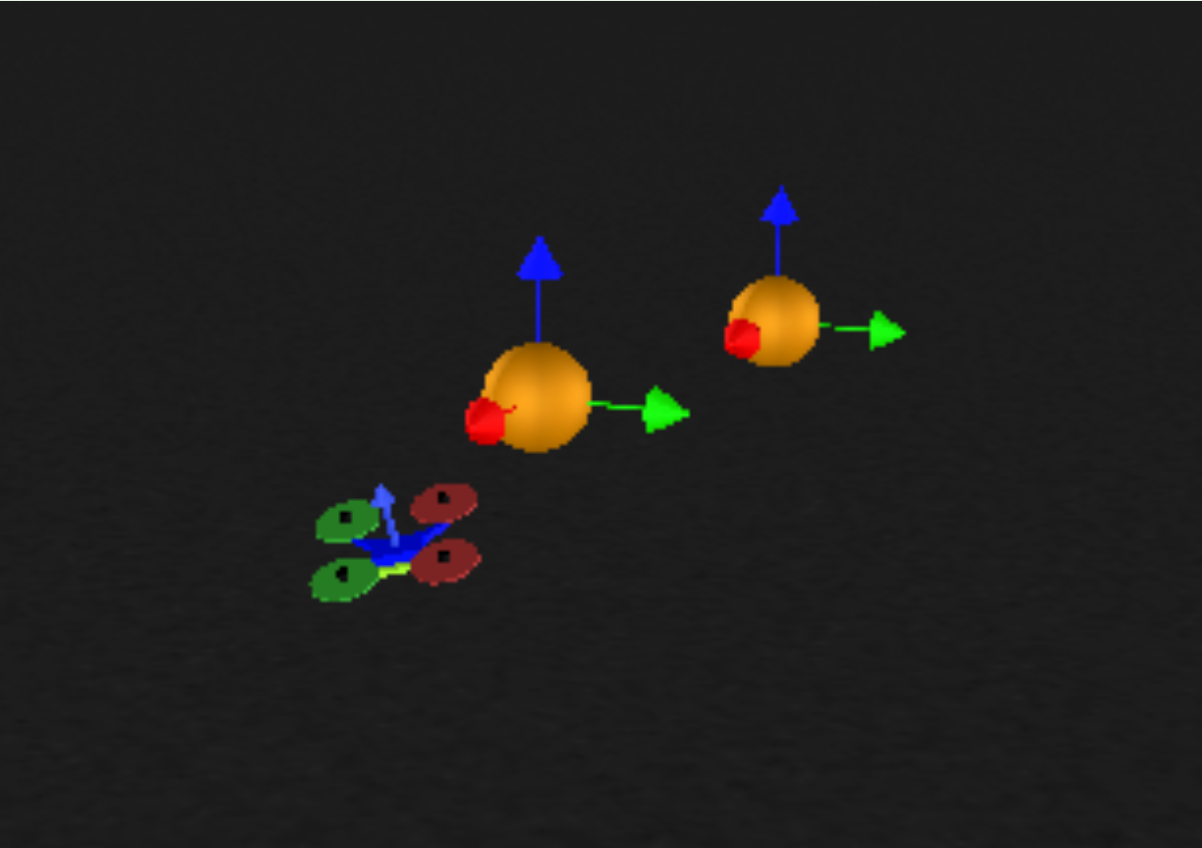}}
        \quad
    \subfloat[Humanoid Project]{\label{fig:robodaycare}

    	\includegraphics[width=3.8cm, height=3.5cm]{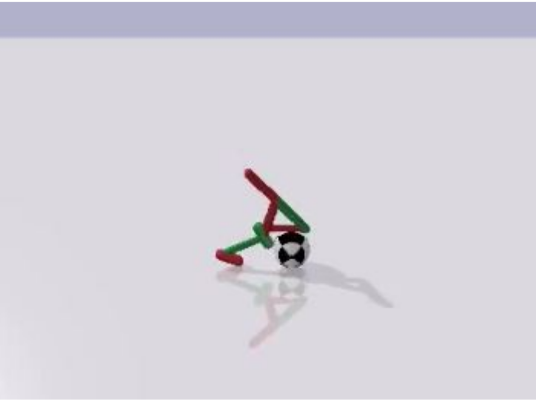}}
    \caption{Example projects simulating and training robotics agents in OpenAI Gym.}
    \label{fig:Environments}
\end{figure}

\subsection{Simulated Interactive Environments}

Motivated by the 2020 Australian Wildfires, a team simulated a wildfire in a forest ecosystem through a Unity video game. The goal is to train an agent, represented by a firefighter dog, to find and save endangered animals from the fire without hurting itself. Using Unity's Machine Learning Library, the team trained the agent using PPO, Behavioral Cloning (BC), and Generative Adversarial Imitation Learning (GAIL) to evaluate how long each agent takes to rescue an animal. 


There were several self-driving student projects in this course. Students developed autonomous driving agents for the games Grand Theft Auto V, Mario kart, Track Mania Nations Forever, TORCS, and Live for Speed racing simulator. Different techniques were applied to each of these projects. Object detection algorithms such as AlexNet, VGG16, YOLOv3, and Hough transformations were implemented to detect the racing track and obstacles within the game and avoid collision with other vehicles. DQN, Imitation Learning, Policy Gradients, and Transfer Learning were experimented with to train the agent to drive.

Another self-navigating team trained an autonomous quadcopter in a simulated 3D environment shown in the left of Figure \ref{fig:Environments}. The team implement a unified quadrotor control policy using PPO, curriculum, and supervised learning to enable the drone going along a specific path. This policy generalized the task as a single learning task, significantly reducing the amount of training needed. In addition to reward shaping, the team tested the model across different track environments, such as a 2D square, 2D square diagonal, and a descending ellipse.

Apart from simulating drones, students also explored on simulated humanoids in robotics seen in the right of Figure \ref{fig:Environments}. Using PyBullet, the agent was trained in varying environments to move a ball to a designated area. Applying reinforcement learning algorithms like PPO and A2C, the team simulated two movements; hopping and walking in different wind environments for each movement. The team defined reward functions based on how long the agent remains alive, how close it is to the target location, and how much movement is taken to achieve the goal. 

\section*{Resources}
For this course, we provide each student with \textdollar 50 Google Cloud Credit (GCP) to be utilized for training Deep Learning algorithms. This sums up to \textdollar 300 for a team of 6 students. In addition to this, students are provided laboratory access to high-end Windows systems. These systems are equipped with NVIDIA GTX 1080 GPUs, 32GB RAM, and Intel i7 7th generation processors. 

This course structure withstood the challenges of transitioning from in-person to hybrid and eventually to fully online modalities throughout the COVID-19 Pandemic. The resistance to risks is mainly credited to the extensive use of Google Cloud services, Github for code version control, Slack and Zoom for instant communication, as well as Piazza and Blackboard for course logistics. The semester-long team project has also provided flexible but adjustable difficulties for both students and instructors.

\section*{Evaluations and Conclusion}

Table ~\ref{tab:evaluation} shows our survey results to evaluate the course. A majority of students gave high ratings for recommending this course to other students, the usefulness of this course for finding an internship or a full-time job, and learning from team projects to get applied machine learning hands-on experiences. The survey results indicate positive feedback for the course.  

From a teaching perspective, we encountered three challenges: At what level should we balance theoretical deep learning and reinforcement learning lecture materials and applied environment demonstrations; How to create an adaptive learning curve for students with varying machine learning backgrounds; and how to form an innovative research pipeline at the graduate school level to facilitate publications. 
Throughout the three semesters, we learned that more visual aids, such as live demonstrations and videos, are needed to increase online engagement as we move into online virtual courses. Weekly project demonstrations in front of the whole class will create a healthy peer effect that increases learning efficacy. Within three semesters, three research conference papers have been published, and more in preparation. From the students' self-proposed projects, we strengthened our belief that gaming as an interdisciplinary research domain can reach other fields, such as robotics, medical diagnosis, human-computer interactions, etc. Games are the testbeds for advancing state-of-the-arts learning algorithms. 
In the future, the class can benefit from state-of-the-arts paper reading sessions and live coding demonstrations to help graduate students build a comprehensive understanding of how a research project is built.

This report summarizes the design of our applied machine learning course for graduate students interested in applying deep learning and reinforcement learning advancements towards gaming. We familiarize students with the current research landscape and improve students' oral and written presentation skills through practical team projects, regardless of the major and machine learning expertise level. Our course can help students open doors for an industry career, aim for publications, or lay the foundations of future products.


\bibliography{eaai21.bib}

\begin{thebibliography}{56}
\providecommand{\natexlab}[1]{#1}
\providecommand{\url}[1]{\texttt{#1}}
\providecommand{\urlprefix}{URL }
\expandafter\ifx\csname urlstyle\endcsname\relax
  \providecommand{\doi}[1]{doi:\discretionary{}{}{}#1}\else
  \providecommand{\doi}{doi:\discretionary{}{}{}\begingroup
  \urlstyle{rm}\Url}\fi

\bibitem[{Astrom(1965)}]{astrom1965optimal}
Astrom, K.~J. 1965.
\newblock Optimal control of Markov decision processes with incomplete state
  estimation.
\newblock \emph{J. Math. Anal. Applic.} 10: 174--205.

\bibitem[{Aytar et~al.(2018)Aytar, Pfaff, Budden, Paine, Wang, and
  de~Freitas}]{aytar2018playing}
Aytar, Y.; Pfaff, T.; Budden, D.; Paine, T.; Wang, Z.; and de~Freitas, N. 2018.
\newblock Playing hard exploration games by watching youtube.
\newblock In \emph{Advances in Neural Information Processing Systems},
  2930--2941.

\bibitem[{Baker et~al.(2019)Baker, Kanitscheider, Markov, Wu, Powell, McGrew,
  and Mordatch}]{baker2019emergent}
Baker, B.; Kanitscheider, I.; Markov, T.; Wu, Y.; Powell, G.; McGrew, B.; and
  Mordatch, I. 2019.
\newblock Emergent tool use from multi-agent autocurricula.
\newblock \emph{arXiv:1909.07528} .

\bibitem[{Baldi(1995)}]{baldi1995gradient}
Baldi, P. 1995.
\newblock Gradient descent learning algorithm overview: A general dynamical
  systems perspective.
\newblock \emph{IEEE Transactions on neural networks} 6(1): 182--195.

\bibitem[{Bard et~al.(2020)Bard, Foerster, Chandar, Burch, Lanctot, Song,
  Parisotto, Dumoulin, Moitra, Hughes et~al.}]{bard2020hanabi}
Bard, N.; Foerster, J.~N.; Chandar, S.; Burch, N.; Lanctot, M.; Song, H.~F.;
  Parisotto, E.; Dumoulin, V.; Moitra, S.; Hughes, E.; et~al. 2020.
\newblock The hanabi challenge: A new frontier for ai research.
\newblock \emph{Artificial Intelligence} 280: 103216.

\bibitem[{Bellemare et~al.(2013)Bellemare, Naddaf, Veness, and
  Bowling}]{bellemare2013arcade}
Bellemare, M.~G.; Naddaf, Y.; Veness, J.; and Bowling, M. 2013.
\newblock The arcade learning environment: An evaluation platform for general
  agents.
\newblock \emph{Journal of Artificial Intelligence Research} 47: 253--279.

\bibitem[{Bellman(1957)}]{bellman1957markovian}
Bellman, R. 1957.
\newblock A Markovian decision process.
\newblock \emph{Journal of mathematics and mechanics} 679--684.

\bibitem[{Berner et~al.(2019)Berner, Brockman, Chan, Cheung, Debiak, Dennison,
  Farhi, Fischer, Hashme, Hesse et~al.}]{berner2019dota}
Berner, C.; Brockman, G.; Chan, B.; Cheung, V.; Debiak, P.; Dennison, C.;
  Farhi, D.; Fischer, Q.; Hashme, S.; Hesse, C.; et~al. 2019.
\newblock Dota 2 with large scale deep reinforcement learning.
\newblock \emph{arXiv preprint arXiv:1912.06680} .

\bibitem[{Brockman et~al.(2016)Brockman, Cheung, Pettersson, Schneider,
  Schulman, Tang, and Zaremba}]{brockman2016openai}
Brockman, G.; Cheung, V.; Pettersson, L.; Schneider, J.; Schulman, J.; Tang,
  J.; and Zaremba, W. 2016.
\newblock Openai gym.
\newblock \emph{arXiv preprint arXiv:1606.01540} .

\bibitem[{C\^ot\'e et~al.(2018)C\^ot\'e, K\'ad\'ar, Yuan, Kybartas, Barnes,
  Fine, Moore, Tao, Hausknecht, Asri, Adada, Tay, and
  Trischler}]{cote18textworld}
C\^ot\'e, M.-A.; K\'ad\'ar, A.; Yuan, X.; Kybartas, B.; Barnes, T.; Fine, E.;
  Moore, J.; Tao, R.~Y.; Hausknecht, M.; Asri, L.~E.; Adada, M.; Tay, W.; and
  Trischler, A. 2018.
\newblock TextWorld: A Learning Environment for Text-based Games.
\newblock \emph{CoRR} abs/1806.11532.

\bibitem[{Fujimoto, Van~Hoof, and Meger(2018)}]{fujimoto2018addressing}
Fujimoto, S.; Van~Hoof, H.; and Meger, D. 2018.
\newblock Addressing function approximation error in actor-critic methods.
\newblock \emph{arXiv preprint arXiv:1802.09477} .

\bibitem[{Girshick(2015)}]{girshick2015fast}
Girshick, R. 2015.
\newblock Fast r-cnn.
\newblock In \emph{Proceedings of the IEEE international conference on computer
  vision}, 1440--1448.

\bibitem[{Girshick et~al.(2014)Girshick, Donahue, Darrell, and
  Malik}]{girshick2014rich}
Girshick, R.; Donahue, J.; Darrell, T.; and Malik, J. 2014.
\newblock Rich feature hierarchies for accurate object detection and semantic
  segmentation.
\newblock In \emph{Proceedings of the IEEE conference on computer vision and
  pattern recognition}, 580--587.

\bibitem[{Goodfellow et~al.(2014)Goodfellow, Pouget-Abadie, Mirza, Xu,
  Warde-Farley, Ozair, Courville, and Bengio}]{goodfellow2014generative}
Goodfellow, I.; Pouget-Abadie, J.; Mirza, M.; Xu, B.; Warde-Farley, D.; Ozair,
  S.; Courville, A.; and Bengio, Y. 2014.
\newblock Generative adversarial nets.
\newblock In \emph{Advances in neural information processing systems},
  2672--2680.

\bibitem[{Graves(2012)}]{graves2012supervised}
Graves, A. 2012.
\newblock Supervised sequence labelling.
\newblock In \emph{Supervised sequence labelling with recurrent neural
  networks}, 5--13. Springer.

\bibitem[{Hecht-Nielsen(1992)}]{hecht1992theory}
Hecht-Nielsen, R. 1992.
\newblock Theory of the backpropagation neural network.
\newblock In \emph{Neural networks for perception}, 65--93. Elsevier.

\bibitem[{Hochreiter(1998)}]{hochreiter1998vanishing}
Hochreiter, S. 1998.
\newblock The vanishing gradient problem during learning recurrent neural nets
  and problem solutions.
\newblock \emph{International Journal of Uncertainty, Fuzziness and
  Knowledge-Based Systems} 6(02): 107--116.

\bibitem[{Hochreiter and Schmidhuber(1997)}]{hochreiter1997long}
Hochreiter, S.; and Schmidhuber, J. 1997.
\newblock Long short-term memory.
\newblock \emph{Neural computation} 9(8): 1735--1780.

\bibitem[{Hu et~al.(2019)Hu, Yarats, Gong, Tian, and
  Lewis}]{hu2019hierarchical}
Hu, H.; Yarats, D.; Gong, Q.; Tian, Y.; and Lewis, M. 2019.
\newblock Hierarchical decision making by generating and following natural
  language instructions.
\newblock In \emph{Advances in neural information processing systems},
  10025--10034.

\bibitem[{Jaderberg et~al.(2019)Jaderberg, Czarnecki, Dunning, Marris, Lever,
  Castaneda, Beattie, Rabinowitz, Morcos, Ruderman et~al.}]{jaderberg2019human}
Jaderberg, M.; Czarnecki, W.~M.; Dunning, I.; Marris, L.; Lever, G.; Castaneda,
  A.~G.; Beattie, C.; Rabinowitz, N.~C.; Morcos, A.~S.; Ruderman, A.; et~al.
  2019.
\newblock Human-level performance in 3D multiplayer games with population-based
  reinforcement learning.
\newblock \emph{Science} 364(6443): 859--865.

\bibitem[{Jin et~al.(2017)Jin, Zhang, Li, Tian, Zhu, and Fang}]{jin2017towards}
Jin, Y.; Zhang, J.; Li, M.; Tian, Y.; Zhu, H.; and Fang, Z. 2017.
\newblock Towards the automatic anime characters creation with generative
  adversarial networks.
\newblock \emph{arXiv preprint arXiv:1708.05509} .

\bibitem[{Johnson et~al.(2016)Johnson, Hofmann, Hutton, and
  Bignell}]{johnson2016malmo}
Johnson, M.; Hofmann, K.; Hutton, T.; and Bignell, D. 2016.
\newblock The Malmo Platform for Artificial Intelligence Experimentation.
\newblock In \emph{IJCAI}, 4246--4247.

\bibitem[{Juliani et~al.(2018)Juliani, Berges, Vckay, Gao, Henry, Mattar, and
  Lange}]{juliani2018unity}
Juliani, A.; Berges, V.-P.; Vckay, E.; Gao, Y.; Henry, H.; Mattar, M.; and
  Lange, D. 2018.
\newblock Unity: A general platform for intelligent agents.
\newblock \emph{arXiv preprint arXiv:1809.02627} .

\bibitem[{{Justesen} et~al.(2019){Justesen}, {Bontrager}, {Togelius}, and
  {Risi}}]{8632747}
{Justesen}, N.; {Bontrager}, P.; {Togelius}, J.; and {Risi}, S. 2019.
\newblock Deep Learning for Video Game Playing.
\newblock \emph{IEEE Transactions on Games} 1--1.
\newblock ISSN 2475-1510.
\newblock \doi{10.1109/TG.2019.2896986}.

\bibitem[{Kakade(2002)}]{kakade2002natural}
Kakade, S.~M. 2002.
\newblock A natural policy gradient.
\newblock In \emph{Advances in neural information processing systems},
  1531--1538.

\bibitem[{Konda and Tsitsiklis(2000)}]{konda2000actor}
Konda, V.~R.; and Tsitsiklis, J.~N. 2000.
\newblock Actor-critic algorithms.
\newblock In \emph{Advances in neural information processing systems},
  1008--1014.

\bibitem[{LeCun et~al.(2012)LeCun, Bottou, Orr, and
  M{\"u}ller}]{lecun2012efficient}
LeCun, Y.~A.; Bottou, L.; Orr, G.~B.; and M{\"u}ller, K.-R. 2012.
\newblock Efficient backprop.
\newblock In \emph{Neural networks: Tricks of the trade}, 9--48. Springer.

\bibitem[{Lee et~al.(2018)Lee, Chen, Yu, and Lai}]{lee2018image}
Lee, S.-J.; Chen, T.; Yu, L.; and Lai, C.-H. 2018.
\newblock Image classification based on the boost convolutional neural network.
\newblock \emph{IEEE Access} 6: 12755--12768.

\bibitem[{Lin et~al.(2017)Lin, Gehring, Khalidov, and
  Synnaeve}]{lin2017stardata}
Lin, Z.; Gehring, J.; Khalidov, V.; and Synnaeve, G. 2017.
\newblock Stardata: A starcraft ai research dataset.
\newblock \emph{arXiv preprint arXiv:1708.02139} .

\bibitem[{Lohokare, Shah, and Zyda(2020)}]{lohokare2020deep}
Lohokare, A.; Shah, A.; and Zyda, M. 2020.
\newblock Deep Learning Bot for League of Legends.
\newblock In \emph{Proceedings of the AAAI Conference on Artificial
  Intelligence and Interactive Digital Entertainment}, volume~16, 322--324.

\bibitem[{Mnih et~al.(2016)Mnih, Badia, Mirza, Graves, Lillicrap, Harley,
  Silver, and Kavukcuoglu}]{mnih2016asynchronous}
Mnih, V.; Badia, A.~P.; Mirza, M.; Graves, A.; Lillicrap, T.; Harley, T.;
  Silver, D.; and Kavukcuoglu, K. 2016.
\newblock Asynchronous methods for deep reinforcement learning.
\newblock In \emph{International conference on machine learning}, 1928--1937.

\bibitem[{Mnih et~al.(2013)Mnih, Kavukcuoglu, Silver, Graves, Antonoglou,
  Wierstra, and Riedmiller}]{mnih2013playing}
Mnih, V.; Kavukcuoglu, K.; Silver, D.; Graves, A.; Antonoglou, I.; Wierstra,
  D.; and Riedmiller, M. 2013.
\newblock Playing atari with deep reinforcement learning.
\newblock \emph{arXiv preprint arXiv:1312.5602} .

\bibitem[{Mnih et~al.(2015)Mnih, Kavukcuoglu, Silver, Rusu, Veness, Bellemare,
  Graves, Riedmiller, Fidjeland, Ostrovski et~al.}]{mnih2015human}
Mnih, V.; Kavukcuoglu, K.; Silver, D.; Rusu, A.~A.; Veness, J.; Bellemare,
  M.~G.; Graves, A.; Riedmiller, M.; Fidjeland, A.~K.; Ostrovski, G.; et~al.
  2015.
\newblock Human-level control through deep reinforcement learning.
\newblock \emph{Nature} 518(7540): 529--533.

\bibitem[{Nair and Hinton(2010)}]{nair2010rectified}
Nair, V.; and Hinton, G.~E. 2010.
\newblock Rectified linear units improve restricted boltzmann machines.
\newblock In \emph{Proceedings of the 27th international conference on machine
  learning (ICML-10)}, 807--814.

\bibitem[{Nichol et~al.(2018)Nichol, Pfau, Hesse, Klimov, and
  Schulman}]{nichol2018retro}
Nichol, A.; Pfau, V.; Hesse, C.; Klimov, O.; and Schulman, J. 2018.
\newblock Gotta Learn Fast: A New Benchmark for Generalization in RL.
\newblock \emph{arXiv preprint arXiv:1804.03720} .

\bibitem[{{Ongsulee}(2017)}]{8259629}
{Ongsulee}, P. 2017.
\newblock Artificial intelligence, machine learning and deep learning.
\newblock In \emph{2017 15th International Conference on ICT and Knowledge
  Engineering (ICT KE)}, 1--6.
\newblock ISSN 2157-0981.
\newblock \doi{10.1109/ICTKE.2017.8259629}.

\bibitem[{Pascanu, Mikolov, and Bengio(2012)}]{pascanu2012understanding}
Pascanu, R.; Mikolov, T.; and Bengio, Y. 2012.
\newblock Understanding the exploding gradient problem.
\newblock \emph{CoRR, abs/1211.5063} 2: 417.

\bibitem[{Redmon et~al.(2016)Redmon, Divvala, Girshick, and
  Farhadi}]{redmon2016you}
Redmon, J.; Divvala, S.; Girshick, R.; and Farhadi, A. 2016.
\newblock You only look once: Unified, real-time object detection.
\newblock In \emph{Proceedings of the IEEE conference on computer vision and
  pattern recognition}, 779--788.

\bibitem[{Redmon and Farhadi(2017)}]{redmon2017yolo9000}
Redmon, J.; and Farhadi, A. 2017.
\newblock YOLO9000: better, faster, stronger.
\newblock In \emph{Proceedings of the IEEE conference on computer vision and
  pattern recognition}, 7263--7271.

\bibitem[{Redmon and Farhadi(2018)}]{redmon2018yolov3}
Redmon, J.; and Farhadi, A. 2018.
\newblock Yolov3: An incremental improvement.
\newblock \emph{arXiv preprint arXiv:1804.02767} .

\bibitem[{Ren et~al.(2015)Ren, He, Girshick, and Sun}]{ren2015faster}
Ren, S.; He, K.; Girshick, R.; and Sun, J. 2015.
\newblock Faster r-cnn: Towards real-time object detection with region proposal
  networks.
\newblock In \emph{Advances in neural information processing systems}, 91--99.

\bibitem[{Risi and Preuss(2020)}]{risi2020chess}
Risi, S.; and Preuss, M. 2020.
\newblock From Chess and Atari to StarCraft and Beyond: How Game AI is Driving
  the World of AI.
\newblock \emph{KI-K{\"u}nstliche Intelligenz} 34(1): 7--17.

\bibitem[{Sapienza et~al.(2018)Sapienza, Zeng, Bessi, Lerman, and
  Ferrara}]{sapienza2018individual}
Sapienza, A.; Zeng, Y.; Bessi, A.; Lerman, K.; and Ferrara, E. 2018.
\newblock Individual performance in team-based online games.
\newblock \emph{Royal Society open science} 5(6): 180329.

\bibitem[{Savva et~al.(2019)Savva, Kadian, Maksymets, Zhao, Wijmans, Jain,
  Straub, Liu, Koltun, Malik, Parikh, and Batra}]{habitat19iccv}
Savva, M.; Kadian, A.; Maksymets, O.; Zhao, Y.; Wijmans, E.; Jain, B.; Straub,
  J.; Liu, J.; Koltun, V.; Malik, J.; Parikh, D.; and Batra, D. 2019.
\newblock Habitat: {A} {P}latform for {E}mbodied {AI} {R}esearch.
\newblock In \emph{Proceedings of the IEEE/CVF International Conference on
  Computer Vision (ICCV)}.

\bibitem[{Schulman et~al.(2017)Schulman, Wolski, Dhariwal, Radford, and
  Klimov}]{schulman2017proximal}
Schulman, J.; Wolski, F.; Dhariwal, P.; Radford, A.; and Klimov, O. 2017.
\newblock Proximal policy optimization algorithms.
\newblock \emph{arXiv preprint arXiv:1707.06347} .

\bibitem[{Shridhar et~al.(2020)Shridhar, Thomason, Gordon, Bisk, Han, Mottaghi,
  Zettlemoyer, and Fox}]{ALFRED20}
Shridhar, M.; Thomason, J.; Gordon, D.; Bisk, Y.; Han, W.; Mottaghi, R.;
  Zettlemoyer, L.; and Fox, D. 2020.
\newblock {ALFRED: A Benchmark for Interpreting Grounded Instructions for
  Everyday Tasks}.
\newblock In \emph{The IEEE Conference on Computer Vision and Pattern
  Recognition (CVPR)}.

\bibitem[{Silver et~al.(2016)Silver, Huang, Maddison, Guez, Sifre, Van
  Den~Driessche, Schrittwieser, Antonoglou, Panneershelvam, Lanctot
  et~al.}]{silver2016mastering}
Silver, D.; Huang, A.; Maddison, C.~J.; Guez, A.; Sifre, L.; Van Den~Driessche,
  G.; Schrittwieser, J.; Antonoglou, I.; Panneershelvam, V.; Lanctot, M.;
  et~al. 2016.
\newblock Mastering the game of Go with deep neural networks and tree search.
\newblock \emph{nature} 529(7587): 484--489.

\bibitem[{Stooke and Abbeel(2018)}]{stooke2018accelerated}
Stooke, A.; and Abbeel, P. 2018.
\newblock Accelerated methods for deep reinforcement learning.
\newblock \emph{arXiv preprint arXiv:1803.02811} .

\bibitem[{Summerville et~al.(2018)Summerville, Snodgrass, Guzdial,
  Holmg{\aa}rd, Hoover, Isaksen, Nealen, and
  Togelius}]{summerville2018procedural}
Summerville, A.; Snodgrass, S.; Guzdial, M.; Holmg{\aa}rd, C.; Hoover, A.~K.;
  Isaksen, A.; Nealen, A.; and Togelius, J. 2018.
\newblock Procedural content generation via machine learning (PCGML).
\newblock \emph{IEEE Transactions on Games} 10(3): 257--270.

\bibitem[{Tian et~al.(2017)Tian, Gong, Shang, Wu, and Zitnick}]{tian2017elf}
Tian, Y.; Gong, Q.; Shang, W.; Wu, Y.; and Zitnick, C.~L. 2017.
\newblock ELF: An Extensive, Lightweight and Flexible Research Platform for
  Real-time Strategy Games.
\newblock \emph{Advances in Neural Information Processing Systems (NIPS)} .

\bibitem[{Vinyals et~al.(2019)Vinyals, Babuschkin, Czarnecki, Mathieu, Dudzik,
  Chung, Choi, Powell, Ewalds, Georgiev et~al.}]{vinyals2019grandmaster}
Vinyals, O.; Babuschkin, I.; Czarnecki, W.~M.; Mathieu, M.; Dudzik, A.; Chung,
  J.; Choi, D.~H.; Powell, R.; Ewalds, T.; Georgiev, P.; et~al. 2019.
\newblock Grandmaster level in StarCraft II using multi-agent reinforcement
  learning.
\newblock \emph{Nature} 575(7782): 350--354.

\bibitem[{Vinyals et~al.(2017)Vinyals, Ewalds, Bartunov, Georgiev, Vezhnevets,
  Yeo, Makhzani, K{\"u}ttler, Agapiou, Schrittwieser
  et~al.}]{vinyals2017starcraft}
Vinyals, O.; Ewalds, T.; Bartunov, S.; Georgiev, P.; Vezhnevets, A.~S.; Yeo,
  M.; Makhzani, A.; K{\"u}ttler, H.; Agapiou, J.; Schrittwieser, J.; et~al.
  2017.
\newblock Starcraft ii: A new challenge for reinforcement learning.
\newblock \emph{arXiv preprint arXiv:1708.04782} .

\bibitem[{Watkins(1989)}]{watkins1989learning}
Watkins, C. J. C.~H. 1989.
\newblock \emph{Learning from delayed rewards.}
\newblock Ph.D. thesis, University of Cambridge.

\bibitem[{Zeng(2020)}]{zeng2020how}
Zeng, Y. 2020.
\newblock How Human Centered AI Will Contribute Towards Intelligent Gaming
  Systems.
\newblock \emph{The Thirty-Fifth AAAI Conference on Artificial Intelligence
  (AAAI-21)} .

\bibitem[{Zeng et~al.(2020)Zeng, Lei, Li, Jiang, Ferrara, and
  Zyda}]{zeng2020learning}
Zeng, Y.; Lei, D.; Li, B.; Jiang, G.; Ferrara, E.; and Zyda, M. 2020.
\newblock Learning to Reason in Round-Based Games: Multi-Task Sequence
  Generation for Purchasing Decision Making in First-Person Shooters.
\newblock In \emph{Proceedings of the AAAI Conference on Artificial
  Intelligence and Interactive Digital Entertainment}, volume~16, 308--314.

\bibitem[{Zeng, Sapienza, and Ferrara(2019)}]{zeng2019influence}
Zeng, Y.; Sapienza, A.; and Ferrara, E. 2019.
\newblock The Influence of Social Ties on Performance in Team-based Online
  Games.
\newblock \emph{IEEE Transactions on Games} .

\end{thebibliography}
\end{document}